\documentclass[12pt]{iopart}

\usepackage{graphicx}
\usepackage{dcolumn}
\usepackage{bm}
\usepackage{float}
\usepackage[font=small,labelfont=bf]{caption}

\begin{document}

\title{Andreev Reflection without Fermi surface alignment in High T$_{c}$-Topological heterostructures}
\author{Parisa Zareapour$^1$, Alex Hayat$^1$, Shu Yang F. Zhao$^1$\footnote{Present Address: Department of Physics, Harvard University, Boston MA USA},Michael Kreshchuk$^1$, Zhijun Xu$^2$, T. S. Liu$^2$\footnote{Present Address: School of Chemical Engineering and Environment, North University of China, China}, G.D. Gu$^2$, Shuang Jia$^3$\footnote{Present Address: International Center for Quantum Materials, School of Physics, Peking University, Beijing 100871, China}, Robert J. Cava$^3$, H.-Y. Yang$^4$, Ying Ran$^4$, and Kenneth S. Burch$^4$}

\address{$^1$Department of Physics and Institute for Optical Sciences, University of Toronto, 60 St George Street, Toronto, Ontario, Canada M5S 1A7}
\address{$^2$Department of Condensed Matter Physics and Materials Science (CMPMS), Brookhaven National Laboratory, Upton, New York 11973, USA}
\address{$^3$Department of Chemistry, Princeton University, Princeton, New Jersey 08544, USA}
\address{$^4$Department of Physics, Boston College, 140 Commonwealth Avenue, Chestnut Hill, MA 02467}
\ead{ks.burch@bc.edu}

\begin{abstract} We address the controversy over the proximity effect between topological materials and high T$_{c}$ superconductors. Junctions are produced between Bi$_{2}$Sr$_{2}$CaCu$_{2}$O$_{8+\delta}$ and materials with different Fermi surfaces (Bi$_{2}$Te$_{3}$ \& graphite). Both cases reveal tunneling spectra consistent with Andreev reflection. This is confirmed by magnetic field that shifts features via the Doppler effect. This is modeled with a single parameter that accounts for tunneling into a screening supercurrent. Thus the tunneling involves Cooper pairs crossing the heterostructure, showing the Fermi surface mis-match does not hinder the ability to form transparent interfaces, which is accounted for by the extended Brillouin zone and different lattice symmetries.
\end{abstract}

\submitto{\NJP}
\maketitle
\section{Introduction}
	Potential novel optical effects\cite{Suemune:2006du,PhysRevB.89.094508} and non-abelian anyons\cite{Oreg:2010gk,Linder:2010hv,Ivanov:2000df,Fu:2009vy,Beenakker:2011tp,Kitaev:2007gb} have reinvigorated interest in the superconducting proximity effect. Various approaches to high temperature superconducting proximity have claimed success\cite{2000PhRvL..85.3708D,Tarutani:1991cr,2004PhRvL..93o7002B,Kabasawa:vj,Sharoni:2004ba,KorenProximityYBCO,weiManganiteYBCO,PhysRevB.67.214511,GolodPRB2013}, including the recent report of a proximity effect between Bi$_{2}$Sr$_{2}$CaCu$_{2}$O$_{8+\delta}$ and Bi$_{2}$Se$_{3}$ or Bi$_{2}$Te$_{3}$,\cite{Zareapour:2012ja,Zareapour2014PRLsub} via the mechanical bonding technique. Using thin films and ARPES, another group has claimed such interfaces result in an s-wave superconductor in the surface states\cite{Wang:2013fpa}. One theoretical study suggested this is due to the mis-match in crystal symmetries\cite{2015PhRvB..91w5143L}, though another finds the d-wave channel is dominant\cite{2016PhRvB..93c5140L}. Two other thin film/ARPES studies suggest the proximity effect is not possible due to the Fermi surface mis-match that suppresses the interface transparency.\cite{Yilmaz:2014jt,Xu:2014cs} 

\begin{figure}
\includegraphics[width=0.98\textwidth]{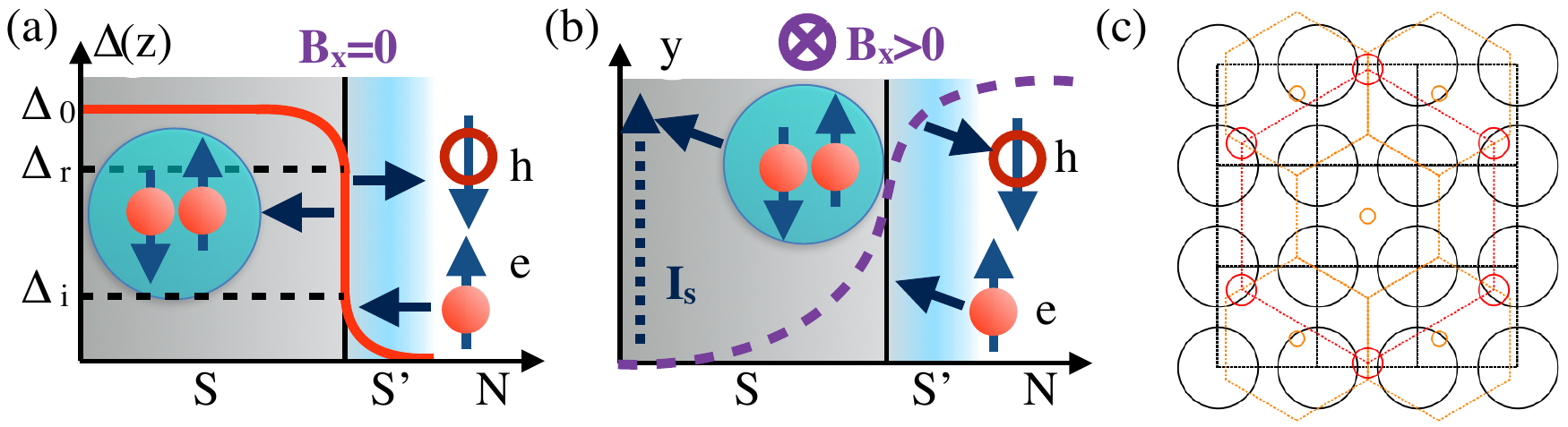}
\caption{\label{fig:Andreev} (a) Andreev reflection process at the superconductor-normal interface without magnetic field. (Red) order parameter, $\Delta_r$ the gap at the superconductor surface, and $\Delta_i$ the induced gap. (b) Affects of applied magnetic field. (c) Fermi surfaces in the extended Brillouin zone. Dotted lines are boundaries of the zones, solid lines are the Fermi surfaces, with black Bi2212, red graphite, and orange Bi$_2$Te$_3$.}
\end{figure}

	We test this hypothesis with tunneling experiments on junctions between Bi$_{2}$Sr$_{2}$CaCu$_{2}$O$_{8+\delta}$ (Bi-2212) and Bi$_{2}$Te$_{3}$ or graphite in magnetic field. These materials are chosen as they form mechanical junctions with different Fermi surfaces. The Bi$_{2}$Te$_{3}$ is hole doped with a Fermi surface close the $\Gamma$ point,\cite{cava2013crystal} whereas graphite is a semimetal with pockets close to the zone boundary (see figure \ref{fig:Andreev}(c))\cite{neto2009electronic}.  By establishing Andreev reflection and proximity, we show that our original efforts which focused on the similar materials (Bi$_{2}$Te$_{3}$ and Bi$_{2}$Se$_{3}$), are not a special case. The ability to obtain Andreev despite the Fermi surface mis-match is explained by considering the extended zone scheme, where the Fermi surfaces meet.

The heterostructure is probed using differential conductance (dI/dV) to look for Andreev reflection, the process where a quasi-particle converts into a Cooper pair and travels from the normal material into the superconductor (see figure \ref{fig:Andreev}(a)). Andreev Reflection is responsible for the proximity effect and is a stringent test of a transparent interface.\cite{Pannetier:2000wr, Deutscher:1991ij, klapwijk2004proximity,Kashiwaya:1996wz} In metals with no attractive potential one expects,\cite{BELZIG19991251} and observes\cite{leSueur:2008kl} a pair amplitude that can result in a minigap whose size depends on device geometry and superconducting phase of the contacts. In confined devices, with two superconductors separated by a normal material, Andreev reflection can produce Andreev Bound States (nABS) inside the normal region.\cite{Tkachov:2013km,leSueur:2008kl,2011NatPh...7..386D,2010NatPh...6..965P,2016PhRvB..93c5307S} The nABS can ultimately produce the super-current produced inside such devices. Since we form planar junctions with one superconductor, and Bi$_{2}$Te$_{3}$/graphite have been shown to superconduct,\cite{zhang2011pressure,hannay1965superconductivity} we do not anticipate nABS playing a crucial role in our results and the induced gaps result from a proximity induced state. As described later, the magnetic field dependance of our spectra is inconsistent with previous measurements of nABS or induced pair amplitudes. 

\section{Experimental Details}
\begin{figure}
\includegraphics[width=\columnwidth]{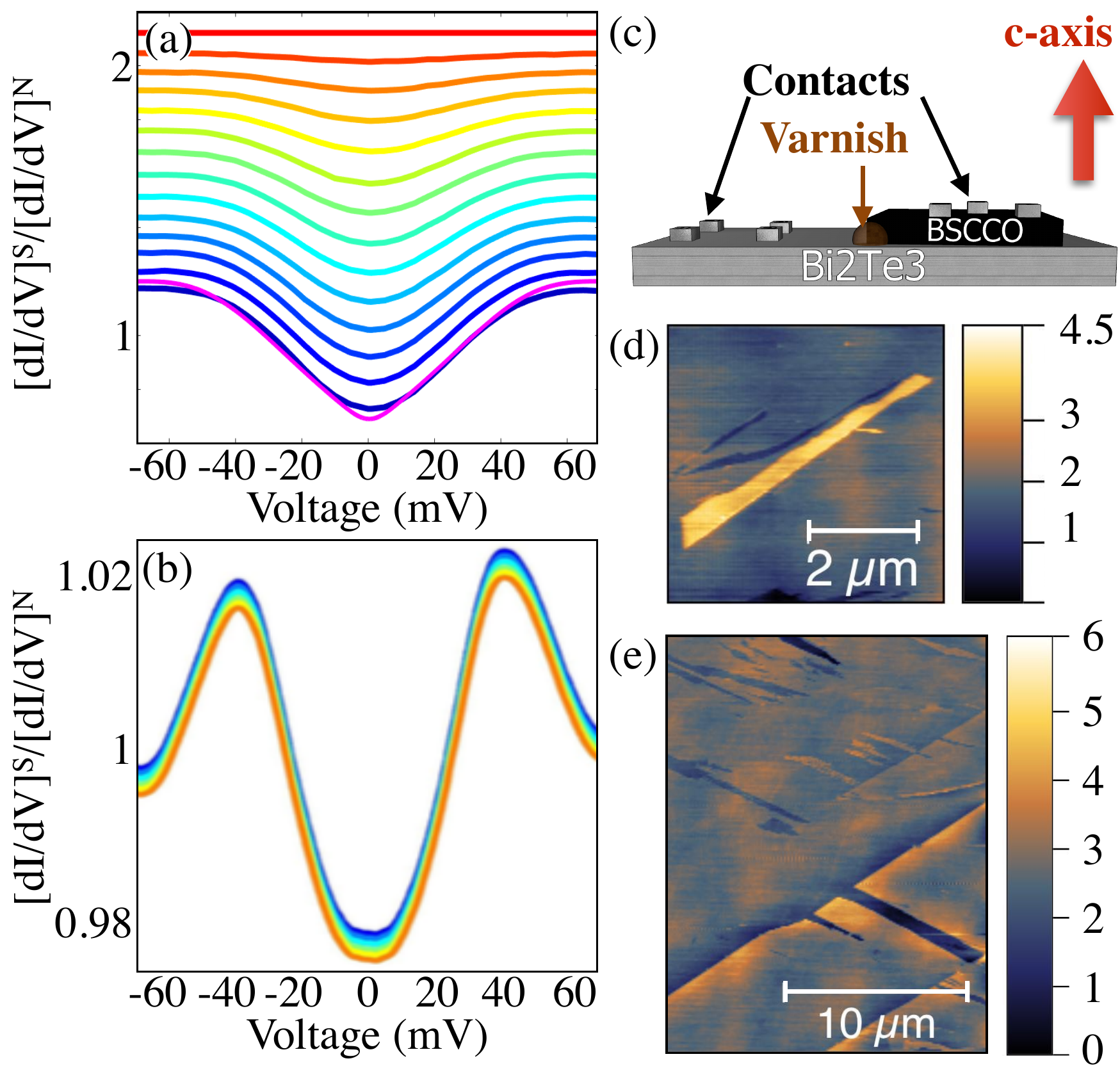}
\caption{\label{fig:dev} (a) dI/dV normalized to its value at T$_{c}$ from a high-barrier Bi$_{2}$Te$_{3}$ junction. Purple line shows a fit with an energy independent lifetime and maximum gap of 40 meV. (b) High barrier junction with offsets due to extra resistance in the graphite upon applying in-plane magnetic field every Tesla from 0 (blue) to 5T (orange). (c) Photo of the device. (d) and (e) AFM from the cleaved surface of Bi$_{2}$Te$_{3}$ revealing atomically flat regions, with some raised mesa's and step edges. Color bar indicates the height in nm.}
\vspace{-4ex}
\end{figure}

	We measured numerous devices with either a high or low barrier. In high barrier devices we observe tunneling spectra consistent with those measured in previous planar junctions, point contact or STM  experiments (see figure \ref{fig:dev}(b)).\cite{2012PhRvB..85u4529S,2003PhyC..387..162G,PhysRevLett.85.1536,Deutscher:2005tm,Aubin:2002go,PhysRevB.62.R14681,DeutscherYBCOField,Bae:2006vt} In addition they are well described by an accepted theory of tunneling into the c-axis of a d-wave superconductor,\cite{Kashiwaya:1996wz,Deutscher:2005tm} with the correct value of the superconducting gap (40 meV). In both Bi$_{2}$Te$_{3}$ and graphite heterostructures with low-barriers,  the conductance is enhanced below T$_{c}$, consistent with Andreev reflection in a proximity induced region. This interpretation is further confirmed by the application of a magnetic field, which generates super-currents. Tunneling into these supercurrents shifts the momentum of the quasi-particles and ultimately their energy. Since this only occurs for the superconducting quasi-particles, the observation of the Doppler effect confirms the presence of Andreev reflection as established in numerous theoretical and experimental works.\cite{Tkachov:2004ed,Rohlfing:2009ju,Tkachov:2005jt,Tkachov:2005et,PhysRevLett.97.027001,WeiDopplerNbSe2,PhysRevB.71.104504,Park:2008bo,Tanaka20031444,PhysRevB.62.R14681,PhysRevB.77.094522,DeutscherYBCOField,PhysRevLett.82.4703} 

	The Doppler effect manifests as a shift in the Andreev features to lower voltages. This is seen in Bi$_{2}$Te$_{3}$ and graphite junctions, confirming they are not due to the normal materials' field dependence. Due to the large size of our devices and the application of the field in the ab-plane, we expect effectively random or no observable shift in ABS. All field dependent spectra are well described by including the Doppler shift in a previously established model of the Andreev reflection for proximity junctions. Furthermore, no hysteresis or splitting of any features is observed, as had been seen in previous measurements of ABS due to the sign change at the 110 surface (dABS).\cite{2003PhyC..387..162G,Deutscher:2005tm} Taken together our results confirm Andreev reflection between Bi$_{2}$Sr$_{2}$CaCu$_{2}$O$_{8+\delta}$ and the normal materials, implying the relative alignment of the Fermi surface is not crucial in these heterostructures. This likely occurs due to the different lattice symmetries that allows the Fermi surfaces to overlap in the extended zone (see figure \ref{fig:Andreev}(c)). 

To form planar junctions between the normal material and Bi-2212 (T$_c$ $\sim$ 90 K) we employed the mechanical bonding method as described in reference \cite{Zareapour:2012ja,Zareapour2014PRLsub}. In an inert glove box both materials are cleaved, then the Bi-2212 was placed on top of the Bi$_2$Te$_3$ or graphite, and GE varnish applied to the Bi-2212 corners (figure \ref{fig:dev}(c)). Four-point transport measurements were performed at various temperatures ranging from 290 K to 10 K. To further clarify the nature of the interface we performed extensive AFM on the cleaved surfaces. The Bi-2212 had extremely flat and large areas, whereas the Bi$_{2}$Te$_{3}$ and graphite produced step edges with atomically flat regions typically tens of microns across (figure \ref{fig:dev}(e)). Some regions showed mesas jutting out (Fig \ref{fig:dev} d). Given the geometry of our devices, this suggests the tunneling is along the c-axis of both materials and occurs at planar junctions formed by these mesas touching the Bi-2212. The dI/dV were highly reproducible regardless of field, temperature, and voltage approach taken. To confirm the dI/dV spectra originate from the junction, for every device we checked different sets of contacts and different combinations of those contacts. Thus our tunneling probes directly the interface between the materials whereas previous studies probed the top surface.\cite{2016PhRvB..93c5140L,Wang:2013fpa,Yilmaz:2014jt,Xu:2014cs} 

	The features we observe might not arise from tunneling given the large contact area. However it is well established that the superconductivity occurs within the Cu-O plane, while the outer layers of the Bi$_{2}$Sr$_{2}$CaCu$_{2}$O$_{8+\delta}$ are insulating Bi-O and Sr-O enabling interlayer josephson junctions and tunneling within a single crystal.\cite{2012PhRvB..85u4529S} Thus we speculate that the Bi-O layer forms the tunnel barrier with the normal material enabling the dI/dV to provide spectra. The large contact area also suggests scattering in either side could occlude the observation of tunneling. This is not the case for our junctions, as shown by the high barrier device (Fig \ref{fig:dev} a). Consistent with previous experiments and the d-wave gap,\cite{2012PhRvB..85u4529S,2003PhyC..387..162G,PhysRevLett.85.1536,Deutscher:2005tm,Aubin:2002go,PhysRevB.62.R14681,DeutscherYBCOField,2012PhRvB..85u4529S,Bae:2006vt} we see a v like shape at low T that gradually fills in as temperature is raised. Furthermore the data are well described by the standard approach to d-wave tunneling in the c-axis producing a proper gap at 40 mV.\cite{Kashiwaya:1996wz,Deutscher:2005tm,PhysRevLett.85.1536}  The fit required a broadening parameter $\Gamma\approx 1\rightarrow4 meV$, much smaller than any of the features observed. This is likely due to the large mobilities of the normal materials used and the c-axis nature of the tunneling, reducing the likelihood of scattering. 
	
	Confirmation that none of the measured features arise from the within the materials is provided by extensive measurements of various contact configurations and application of magnetic field. Four-point and two-point measurements on only one side of the junction were independent of voltage and all reported features were only observed when current and voltage were measured across the interface. Two-point measurements on one side of the junction resulted in resistances $1\rightarrow 10\Omega$ at low T. Swapping the current and voltage leads on the normal material only produced a slight, voltage independent offset of a few $\Omega$, while high barrier devices had resistances $> 1k\Omega$ and low barrier junctions had $dV/dI \approx 100 \Omega$, confirming the voltage drop primarily occurs at the interface. Further confirmation of the voltage independence of the materials contribution is shown in figure \ref{fig:dev}(b), where we measured the response from another high barrier junction between graphite and Bi-2212 at 10K in various applied magnetic fields (similar results were observed with Bi$_{2}$Te$_{3}$). Consistent with STM measurements we find no field dependence of the spectra,\cite{PhysRevLett.85.1536} and a slight, voltage independent offset due to the graphite magneto-resistance.
	
\section{Model}
To understand the dI/dV from the low-barrier junctions, we review what is expected, observed in proximity devices by tunneling\cite{Zareapour:2012ja,Zareapour2014PRLsub,Wolf:1982uv,PCvanSon:2011wx,1982PhRvB..25.4515B,Anonymous:-hBn7hu9,klapwijk2004proximity} and confirmed with ARPES\cite{xu2014momentum}. The superconducting order parameter is induced into the normal material by the conversion of a quasi-particle current into a supercurrent via Andreev reflection. This involves an electron crossing the interface by forming a Cooper pair resulting in a doubling of conductance (figure \ref{fig:Andreev} (a)). In less transparent interfaces, the conductance at zero bias is smaller than two and the shape at finite voltage is altered\cite{1982PhRvB..25.4515B,Kashiwaya:1996wz}. The zero-bias feature's width reflects the full gap of the superconductor. However, Andreev reflection can induce superconductivity in the normal material and thus reduce the superconductor's gap at the surface. Thus we define an induced ($\Delta_i$) and reduced gap ($\Delta_r$) at the interface (figure \ref{fig:Andreev}(b)). These produce Andreev features in the dI/dV that are tell-tale signs of the proximity effect\cite{1982PhRvB..25.4515B,Zareapour:2012ja,Zareapour2014PRLsub,Anonymous:-hBn7hu9,klapwijk2004proximity,xu2014momentum,Wolf:1982uv,PCvanSon:2011wx}.  

The observed features are reproduced by modifying a standard approach to tunneling into d-wave superconductors.\cite{1982PhRvB..25.4515B,Kashiwaya:1996wz,HighTcProxThry} Specifically, the differential conductance below T$_c$ [dI/dV]$_S$, divided by the normal state conductance [dI/dV]$_N$ is given by the half-sphere integration over solid angle $\Omega$:

\begin{eqnarray}
\sigma (E) = \frac{\int {d\Omega cos \theta_N \sigma _S(E)}}{\int {d\Omega\sigma_N cos \theta_N }}
\end{eqnarray}

where E is the quasiparticle energy and $\theta$ is the incidence angle (relative to the interface normal) in the normal material, $\sigma_N$ is the conductance from normal to normal material with the same geometry, and

\begin{eqnarray}
\sigma_S = \frac{\sigma_N(1+\sigma_N |k_+|^2 + (\sigma_N -1)|k_-k_+|^2)}{1 + (\sigma_N -1)|k_-k_+|^2exp(i\phi_--i\phi_+)}
\end{eqnarray}
 
where $k_\pm = \frac{E - \sqrt {|E^2|-|\Delta_\pm^2|}}{|\Delta_\pm|}$ and $\Delta_\pm = |{\Delta_\pm}|exp(i\phi_\pm)$, electron-like and hole-like quasiparticle effective pair potentials with the corresponding phases $i\phi_\pm$. In the case of c-axis tunneling, the hole-like and the electron-like quasiparticles transmitted into the superconductor experience the same effective pair potentials, which have similar dependence on the azimuthal angle $\alpha$ in the AB-plane $\Delta_+ = \Delta_- = \Delta_0cos(2\alpha)$. Scattering-induced energy broadening ($\Gamma$) is included in the calculation by adding an imaginary term to the energy of the quasi-particles. The real part of the resulting $\sigma_S$ then gives the differential conductance (G) with broadening from the normal material included via the $\Gamma$ term ($G (\Delta,\Gamma,V) = Re[\sigma]$, where V is the applied bias).\cite{1978PhRvL..41.1509D,Pekola:2010iaa}

\begin{figure}
\includegraphics[width=0.8\textwidth]{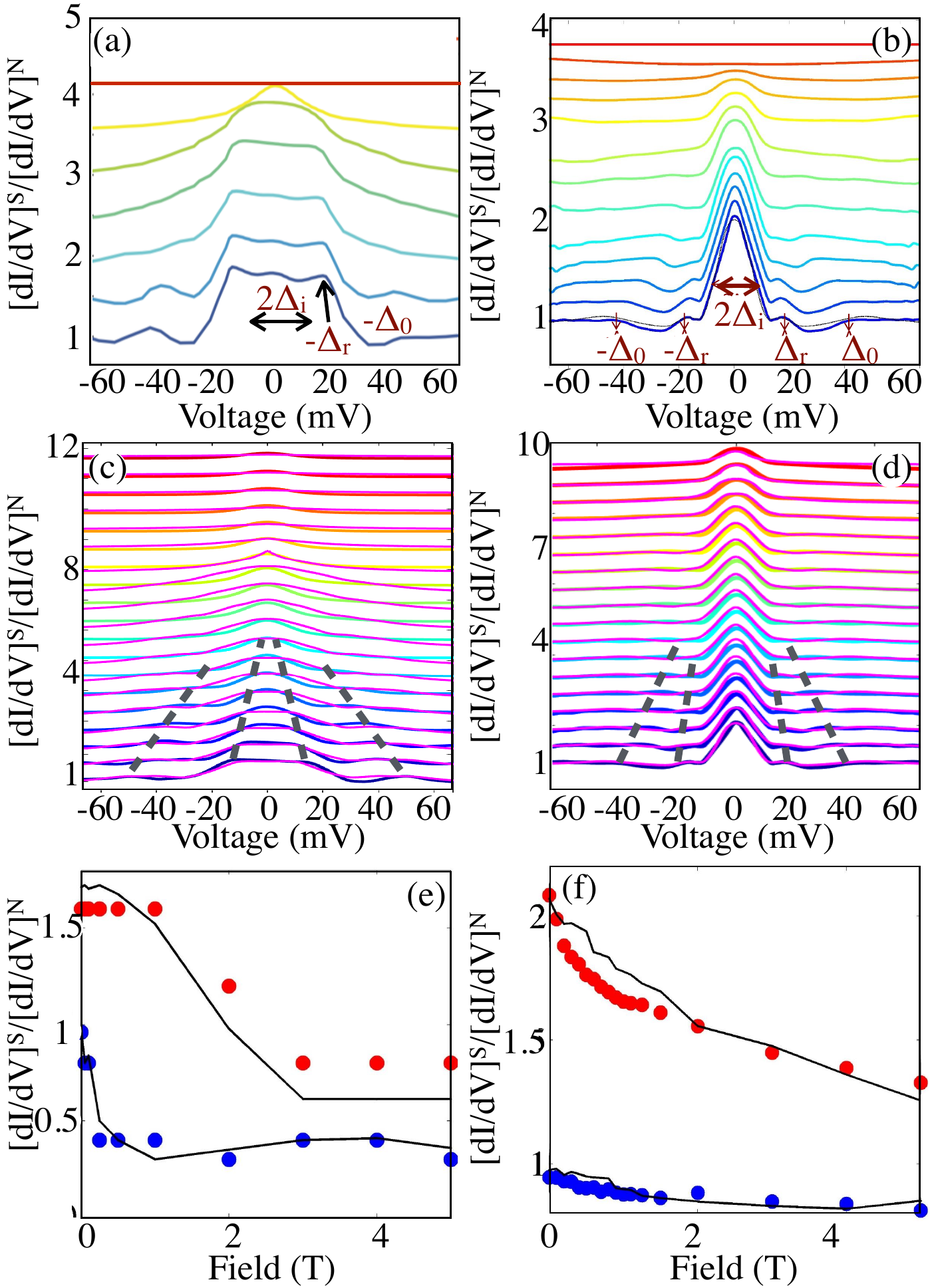}
\caption{\label{fig:compare}  (a,c,e) Bi-2212/ Bi$_{2}$Te$_{3}$ and (b,d,f) Bi-2212/graphite.  (a-b) Offset, normalized dI/dV of  Bi$_{2}$Te$_{3}$ (75K, 60K, 50K, 40K, 30K and 20K), and graphite data every 10 K from T$_{c}$ to 70 K,  then every 5K till 25K, and finally 11K. Three Peaks due to Andreev reflection into induced ($\Delta_{i}$), reduced ($\Delta_{r}$) and full gaps ($\Delta_{0}$). (c-d) dI/dV in parallel magnetic field, with the fits shown in purple. For Bi$_{2}$Te$_{3}$ the fields are (0.05, 0.1, 0.25, 0.5, 1, 2, 3, 4, 5, 6, 7, and 7.5 T), while for graphite they are (every 0.1 T till 1T, then 1.1, 1.25, 1.5, 2, 3, 4 and 5T) Features shift due to the Doppler effect. (e-f) Dots show the measured data at 43 mV (blue) and 0 mV (red) versus field. The black lines are fits with Doppler.}
\vspace{-4ex}
\end{figure}

The total Andreev reflection spectrum at zero magnetic field is obtained by calculating the reflection and the transmission in the proximity region, followed by reflection at the interface between the two materials\cite{Strijkers:2001ii}. Incoming quasiparticles with energies smaller than $\Delta_i$, Andreev-reflect at the first interface. This gives rise to a $G(\Delta_{induced},\Gamma,V)$, which is expected to consist of a central peak with a width typically much smaller than the bulk gap of Bi-2212, corresponding to the induced gap in the normal material ($G(\Delta_{induced},\Gamma,V)$). Quasiparticles with higher energies do not Andreev-reflect off the first interface, and instead transmit as normal particles. The transmission rate is 2-$G(\Delta_{induced},\Gamma,V)$, where the 2 accounts for the fact that Andreev reflection involves a charge of -2e due to the conversion of two normal quasi-particles into a cooper pair (figure \ref{fig:Andreev} B). Quasiparticles with energies between $\Delta_i$ and $\Delta_r$ Andreev-reflect at the second interface and give rise to a term $G(\Delta_{reduced},\Gamma,V)$. We expect $G(\Delta_{reduced},\Gamma,V)$ to consist of peaks at an energy smaller than the bulk gap of Bi-2212 ($\Delta_{reduced}$), due to the suppression of superconductivity at the interface. Lastly an additional term is included, $G(\Delta_{bulk gap},\Gamma,V)$, which can arise due to the inhomogeneity in the tunnel junction (a few high-barrier junctions in parallel with the low-barrier proximity junction). The total dI/dV is calculated using $G_{total} = f_{1}\times[G(\Delta_{induced},\Gamma,V) + (2 - G(\Delta_{induced},\Gamma,V))G(\Delta_{reduced},\Gamma,V)]+ f_{2}\times G(\Delta_{bulk gap},\Gamma,V)$. The parameters $f_{1}$ and $f_{2}$ account for the different relative areas of the proximity and high barrier junctions. Specifically the two types of junctions are in parallel such that their total conductance is equal to their relative volume fractions ($f_{1,2}$) times their intrinsic conductivity. The calculated spectra in this model with the barrier strength (Z), scattering-induced energy broadening ($\Gamma$), and the superconducting gap ($\Delta$) used as fit parameters, show excellent agreement with the experimental conductance measurements (purple lines in figure  \ref{fig:dev} (a), \ref{fig:compare} (c) \& (d)).

\section{Results and Discussion}

The temperature dependent, dI/dV of Bi-2212/Bi$_{2}$Te$_{3}$ junctions are shown in Fig \ref{fig:compare} a. The dI/dV, when normalized to T$_{c}$, reveals a zero bias peak that emerges just below T$_{c}$ and eventually evolves into three features at low temperature. The first is an Andreev reflection peak near zero bias, whose width is much smaller then the full gap of Bi-2212 (labeled $\Delta_{i}$). This feature could be a dABS.\cite{2003PhyC..387..162G,Deutscher:2005tm,Kashiwaya:1996wz} However, at low T, the feature reaches an amplitude of nearly twice the normal state conductance, consistent with standard Andreev reflection. Furthermore, ab-plane tunneling only reveals a narrow peak at zero-bias and at the full gap at Bi-2212 ($\Delta_{0}$). The full gap is also seen in our data,  it's temperature dependence matches well that observed in high barrier devices (Fig \ref{fig:dev}) and established trends for the cuprates. Another peak in our data ($\Delta_{r}$) around 20 meV also approaches 2. Taken together these three peaks are consistent with the proximity effect (Fig \ref{fig:Andreev} a). Specifically we expect perfect Andreev reflection since there is effectively no barrier between the normal material and the induced superconductor, while the inverse proximity effect reduces the size of the gap at the interface resulting in a peak at the reduced ($\Delta_{r}$) gap, and at the full gap ($\Delta_{0}$). We find similar spectra and temperature dependence from a low-barrier device with graphite as the normal material (figure \ref{fig:compare}(b)). The fact that we observe a zero-bias feature of height nearly 2 and the full gap at 40 meV, adds confidence that these features arise from Andreev reflection and do not require the normal material's Fermi surface to directly overlap the superconductor's. 

To confirm these features result from Andreev reflection, we use an in-plane magnetic field $B=\nabla\times A(r)$, which dramatically affects the ABS but causes little magneto-resistance in the normal materials. The superconducting order parameter will acquire an inhomogeneous phase ($\phi(r)=-2\frac{\pi}{\phi_{0}}\int A(r')dr'$), with $\phi_0$ the flux quantum. This produces a diamagnetic screening current where the Cooper pairs acquire a momentum $\hbar k_{s}=\nabla \gamma$. Since Andreev reflection involves tunneling into this supercurrent, they are Doppler shifted by: $E_{D}=-\frac{\hbar^2k_{\perp}k_{s}}{2m_{e}}$, where $k_{\perp}$ is the transverse wavenumber. Thus magnetic fields shift Andreev reflection features by $\Delta E= -v_{F}  P_{S} sin\theta$ (where v$_F$, P$_S$, and $\theta$ are the normal material Fermi velocity, Cooper pair momentum, and the angle between the electron trajectory and magnetic field). The superfluid momentum ($P_S$) is linearly proportional to $B$ and includes a geometric factor whose exact size is difficult to estimate in proximity devices.\cite{Tkachov:2004ed,Rohlfing:2009ju,Tkachov:2005jt,Tkachov:2005et,2003PhyC..387..162G,Deutscher:2005tm,PhysRevB.71.104504,Park:2008bo,PhysRevB.62.R14681,PhysRevB.77.094522,DeutscherYBCOField} Thus for features involving tunneling of cooper pairs, we expect a Doppler shift: $\Delta E= D\times B$, where D is a constant. 

In studies of dABS in Bi-2212, magnetic field split and/or suppressed the zero bias peak due to the Doppler effect, except when the field was aligned in the ab-plane.\cite{2003PhyC..387..162G,Deutscher:2005tm} Alternatively, in S/N/S or N/S/N one can observe nABS due to multiple Andreev reflections between the interfaces. These nABS, which can play a crucial role in the proximity effect and supercurrents, have an energy defined by the phase difference across the confined region, $\Delta_{0}$ and the transparency of the barrier.\cite{Tkachov:2013km} In a confined structure with an induced pair amplitude, the resulting minigap was reduced upon by a \textit{perpendicular} magnetic field,\cite{leSueur:2008kl} as expected from theory\cite{BELZIG19991251} due to $\phi$ between the superconducting contacts.

The situation in our devices is quite different than typically observed in low T$_{c}$ structures. Due to the large H$_{c2}\approx 90T$, we do not expect significant shifts in $\Delta_{0}$ for the fields applied here ($\leq 7.5 T$).\cite{PhysRevLett.86.5763} However, if nABS emerged due to the small mesas seen in AFM, then their energy of will be periodic in the magnetic field as the phase winds through 2$\pi$.\cite{Tkachov:2013km,2011NatPh...7..386D,2010NatPh...6..965P,2016PhRvB..93c5307S} Using the cross-sectional area of our Bi$_{2}$Sr$_{2}$CaCu$_{2}$O$_{8+\delta}$ ($\approx 10^{-7}~m^{2}$) or accounting for the small size of the mesas, we find the features would reach zero energy at $B\approx 10^{-5}~T$. Thus if our features arise from nABS, their shifts with field would essentially be random for the sizes of the fields we apply. Furthermore nABS should only appear below the induced gap. Hence the feature at 40 meV should be tunneling into the full gap of Bi$_{2}$Sr$_{2}$CaCu$_{2}$O$_{8+\delta}$, whereas the features at lower energy could be nABS. As such while the field may tune the zero-bias peak and the peak at $\Delta_{r}$, the full gap will not be affected. 

	In figure \ref{fig:compare}(c\&d) we show the dI/dV spectra at 6.5 K with magnetic field. The size of the dI/dV at zero bias and 43 mV is shown in figure\ref{fig:compare}(e\&f), revealing a systematic suppression of the spectra, suggesting this may be the result of dABS. However the full gap, reduced gap and width of the induced gap feature all move to zero bias upon applying the magnetic field. This provides strong evidence against ABS as the full gap ($\Delta_{0}$) should be field independent, as seen in high barrier junctions (figure \ref{fig:dev} (b)). Thus we attribute the magnetic field induced changes to the Doppler effect. 
	
	This is confirmed by including the Doppler effect in our calculation of the c-axis conductance spectra using the formalism developed for anisotropic superconductors\cite{Kashiwaya:1996wz}. As described in reference \cite{Zareapour:2012ja,Zareapour2014PRLsub}, we modified this formalism to include contributions from the induced, reduced and fully gapped regions. A proper theoretical approach would self-consistently calculate the gap and account for potential multiple reflections due to the gradual change in the gap. However since we do not investigate confined structures, and our minimal model captures our results, we believe it is appropriate to investigate the effects of magnetic field. Once the zero-field spectra are captured by our model, we follow the established procedure for the Doppler effect by calculating the Andreev reflection probability $a(\tilde{E})$ with $\tilde{E}=E+\Delta E$. The entire field dependence is reproduced using only the Doppler factor D. The resulting dI/dV and their values at fixed bias are shown in Figs. \ref{fig:compare}. The excellent agreement, despite only one free parameter, confirms the field dependence is governed by Andreev reflection and the Doppler effect. 

\begin{figure}
\includegraphics[width=0.5\textwidth]{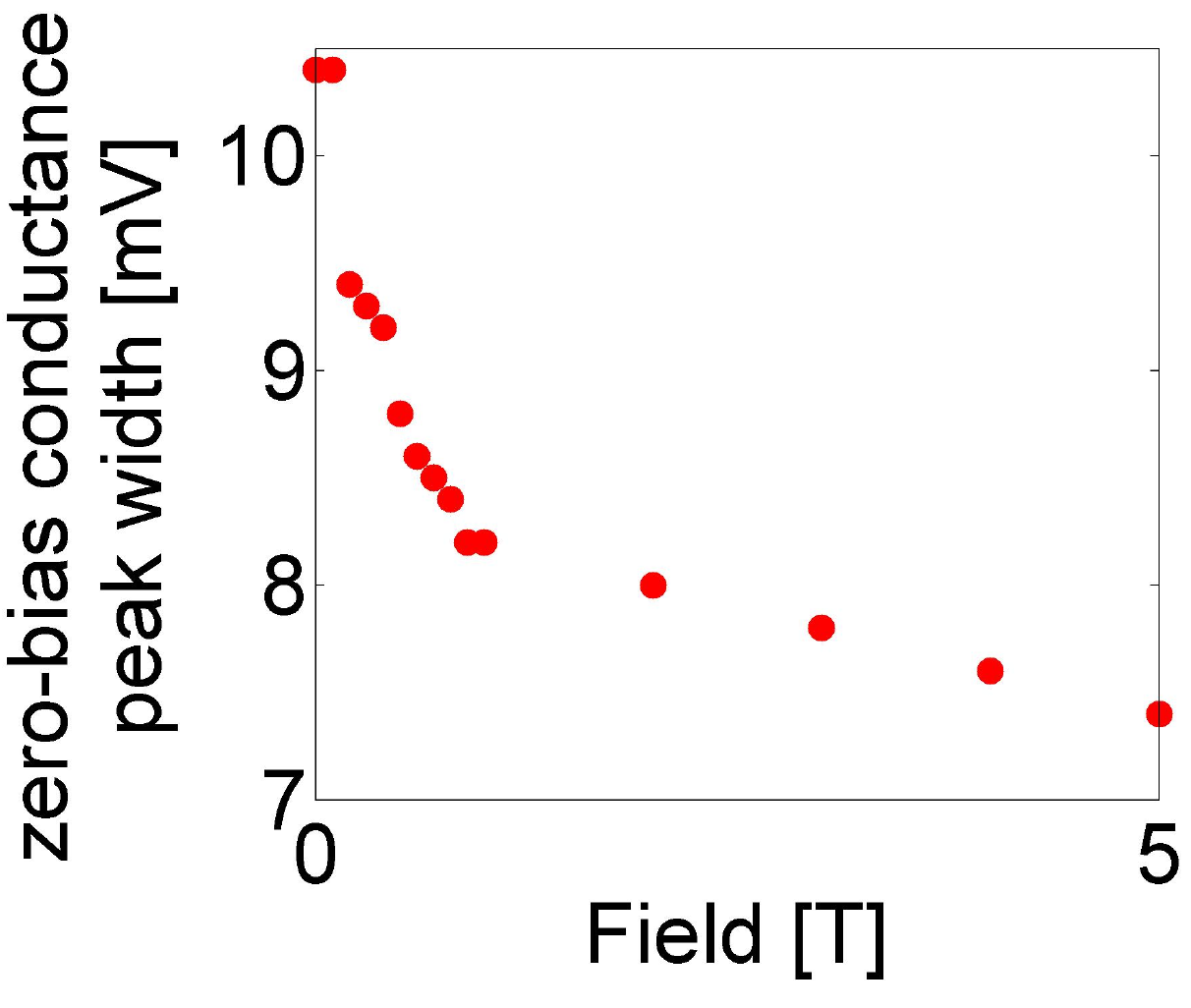}
\caption{\label{fig:supp2}  The width of the induced gap in graphite at 11 K at various applied fields.}
\end{figure}

Before closing, let us discuss some alternate possibilities. For example, Andreev bound states formed at  the 110 interface between a d-wave superconductor/normal metal can create a peak at zero bias\cite{2003PhyC..387..162G,Aubin:2002go,Kashiwaya:1996wz}. However, the central peak in our data is very different from an Andreev bound state. We observe a reduced gap as well as the central gap, in spite of the Andreev bound state only showing up as a zero-bias peak. Lastly, the Andreev bound states are expected to split by the application of magnetic field, while the central peak (the induced gap) in our data, not only does not split, but the width decreases with field. (Fig \ref{fig:supp2} A) \cite{2003PhyC..387..162G,Aubin:2002go}

Other bound states such as geometrical resonances (McMillan-Rowel and Tomasch oscillations)\cite{Shkedy:2004kca,Chang:2004fl} can create peaks in the differential conductance spectra. However, these peaks emerge at certain voltage positions in the Andreev spectra. Tomasch oscillations are due to resonances in the superconductor and create resonance features in dI/dV at voltages given by: $eV_n = \sqrt{(2\Delta)^2 + (\frac{nhv_{fS}}{2d_S})^2}$, with $\Delta$ being the superconducting energy gap, $v_{fS}$ being the Fermi velocity in the superconductor, $d_S$ being the thickness of the superconductor, and n being the dip number). McMillan-Rowell oscillations occur due to geometrical resonances in the normal material and the voltages of the oscillatory features are linear with n (${\Delta}V=\frac{hv_{fN}}{4ed_N}$, with $v_{fN}$ being the Fermi velocity in the normal material and $d_N$ being the thickness of the normal layer at which the reflections occur. Neither of these oscillations agree with the peaks we observe in our data. Furthermore, these oscillations typically create features at finite bias, in contrast to our data where we observe a zero-bias conductance peak. \cite{Shkedy:2004kca} Nonetheless, the magnetic field dependence of McMillan-Rowell and Tomasch oscillations are different from our data. These bound states are expected and seen to split as well as shift by the application of magnetic field (A. Shailos et al., EPL, 79, 57008(2007)). 

A proximity effect along the c-axis of Bi-2212 seems surprising given the short coherence length. However the proximity effect in the cuprates is governed by their low diffusion coefficient and small density of states.\cite{Deutscher:1991ij} Thus the Andreev reflection and proximity effect observed here could result from the small density of states of Bi$_{2}$Te$_{3}$/graphite, and their poor c-axis transport. What about the mis-match in the Fermi surfaces between Bi-2212 and the normal materials? Since the normal materials' lattice symmetries are quite distinct from Bi-2212, the Fermi surfaces touch in the extended Brillouin zone.\cite{neto2009electronic,cava2013crystal,yamamoto1990rietveld} This is shown in figure \ref{fig:Andreev}(c), where the Fermi level of the hole doped Bi$_2$Te$_3$ is around the Dirac point\cite{hor2010development}, and the graphite is of ABAB stacking.\cite{guinea2006electronic} This argument holds for a wide range of Fermi levels.
	
	Strong similarities between Bi$_{2}$Te$_{3}$ and graphite devices may suggest our results are intrinsic to Bi-2212. For example, the junctions could lead to strain that produces mechanical breaks. Indeed, some devices had sharp features in the dI/dV as seen in point contact.\cite{PhysRevLett.107.217001} These result from reaching the critical current in the inhomogeneous superconductor. However such devices did not show the features reported here, and the sharp peaks were suppressed much faster in applied magnetic field then expected from Doppler. Furthermore the Andreev features only appear in measurements performed across the interface. 

\ack
The work at the University of Toronto was supported by the Natural Sciences and Engineering Research Council of Canada, the Canadian Foundation for Innovation, and the Ontario Ministry for Innovation. KSB acknowledges support from the National Science Foundation (grant DMR-1410846). The work at Brookhaven National Laboratory (BNL) was supported by DOE under Contract No. DE-AC02-98CH10886. The crystal growth at Princeton was supported by the US National Science Foundation, grant number DMR-0819860.

\bibliographystyle{iopart-num}
\providecommand{\newblock}{}

\end{document}